\documentclass[conference]{IEEEtran}
\IEEEoverridecommandlockouts
\usepackage{cite}
\usepackage{hyperref}
\usepackage{amsmath,amssymb,amsfonts}
\usepackage{algorithmic}
\usepackage{graphicx}
\usepackage{textcomp}
\usepackage{xcolor}
\usepackage{booktabs} 
\usepackage{multirow}
\usepackage{color}
\usepackage{colortbl}
\usepackage{float}

\definecolor{hred}{RGB}{247,202,197}
\definecolor{horange}{RGB}{242,217,198}
\definecolor{hyellow}{RGB}{247,225,147}
\definecolor{hlgreen}{RGB}{200,207,141}
\definecolor{hdgreen}{RGB}{172,215,156}
\definecolor{hlblue}{RGB}{181,228,227}
\definecolor{hdblue}{RGB}{188,220,246}
\definecolor{hpurple}{RGB}{219,218,246}
\definecolor{tablebest}{RGB}{192,192,192}
\definecolor{myred}{RGB}{239,71,111}
\definecolor{myblue}{RGB}{17,138,178}
\definecolor{mygreen}{RGB}{6,214,160}

\definecolor{incred}{RGB}{10,170,255}
\definecolor{decgreen}{RGB}{6,214,160}

\definecolor{results_red}{RGB}{6,214,160}
\definecolor{results_blue}{RGB}{255,50,80}
\newcommand{\inc}[2]{\cellcolor{incred!#1}#2}
\newcommand{\dec}[2]{\cellcolor{results_red!#1}#2}

\newcommand{\increase}{\colorbox{incred}{\textsc{Blue}}}
\newcommand{\decrease}{\colorbox{decgreen}{\textsc{green}}}

\newcolumntype{M}[1]{>{\centering\arraybackslash}m{#1}}

\usepackage{chemformula} 

\newcommand{\greenlight}[1]{\textcolor{purple}{\bf\small #1}}

\def\BibTeX{{\rm B\kern-.05em{\sc i\kern-.025em b}\kern-.08em
    T\kern-.1667em\lower.7ex\hbox{E}\kern-.125emX}}
\begin{document}

\title{Combining Financial Data and News Articles for Stock Price Movement Prediction Using Large Language Models\\
}

\author{\IEEEauthorblockN{Ali Elahi$^\clubsuit$ Fatemeh Taghvaei$^\diamondsuit$} \\
\IEEEauthorblockA{\textit{$\clubsuit$ Department of Computer Science University of Illinois Chicago}\\
\textit{$\diamondsuit$ Department of Electrical and Computer Engineering University of Illinois Chicago}\\ \\
\{aelahi6, ftaghv2\}@uic.edu}}


\maketitle

\begin{abstract}
    Predicting financial markets and stock price movements requires analyzing a company's performance, historic price movements, industry-specific events alongside the influence of human factors such as social media and press coverage. We assume that financial reports (such as income statements, balance sheets, and cash flow statements), historical price data, and recent news articles can collectively represent aforementioned factors.

    We combine financial data in tabular format with textual news articles and employ pre-trained Large Language Models (LLMs) to predict market movements. Recent research in LLMs has demonstrated that they are able to perform both tabular and text classification tasks, making them our primary model to classify the multi-modal data. We utilize retrieval augmentation techniques to retrieve and attach relevant chunks of news articles to financial metrics related to a company and prompt the LLMs in zero, two, and four-shot settings. Our dataset contains news articles collected from different sources, historic stock price, and financial report data for 20 companies with the highest trading volume across different industries in the stock market. We utilized recently released language models for our LLM-based classifier, including GPT- 3 and 4, and LLaMA- 2 and 3 models.

    We introduce an LLM-based classifier capable of performing classification tasks using combination of tabular (structured) and textual (unstructured) data. By using this model, we predicted the movement of a given stock's price in our dataset with a weighted F1-score of 58.5\% and 59.1\%  and Matthews Correlation Coefficient of  0.175 for both 3-month and 6-month periods.
\end{abstract}


\section{Introduction}

With the advancement of language models and their extensive use in the finance industry \cite{Li2023-dp}, these models are now applied in finance for various purposes. These include text classification tasks \cite{Loukas2023-ex, Vamvourellis2022-sj}, such as semantic analysis of news \footnote{Referring to classifying news as positive or negative} \cite{Zhang2023-yi}. Investors also use language models like ChatGPT to enhance their financial understanding, and there are studies on financial chatbots and fine-tuning LLMs for financial advisement \cite{Wang2023-vp, Lakkaraju2023-rd}.

Predicting market trends and making informed investment decisions has remained an important task in finance. One use of LLMs in this field is predicting if the price is going up or down in the future, known as Price Movement Prediction \cite{Wang2024-cn, Tong2024-qm, Weng2017-aj}. For this purpose, different studies utilize various sources of data to predict price movements.

The performance of a stock in the future is typically influenced by all available information in the current state \cite{MEH}. Numerous factors, including the company's performance, industry dynamics, human responses, social media, news, and broader political and economic shifts, can impact market behavior. We aim to gather as much information as possible to represent these factors for the stock price movement task. This information can be in tabular or textual format; therefore, we aim to design an LLM-based model that can perform stock price movement classification using data in different formats. We also believe that prompting LLM with different sources of information can prevent hallucination.

Major trend in the natural language processing research has shifted to finding the right setup to utilize pre-trained LLMs to solve problems instead of training new models. Additionally,  fine-tuning language models requires significant time, computational resources, and a large dataset, therefore we will conduct this analysis in a zero to four-shot setting. 

The following paragraphs will include our sources of information:

\paragraph{Financial Factors} Every publicly traded company in the United States is mandated by the Securities and Exchange Commission (SEC) to disclose its financial information at the end of each quarter, and among these regulatory filings, the 10-K report stands out as a comprehensive document detailing the company's financial performance, risks, and operational insights, reported on a quarterly basis. To include the company's performance, we included the data available in financial reports (a.k.a. 10-K files) in our analysis.

\paragraph{Historical Price} The trend in historical price changes for a company can indicate potential future movements, making historical price data important for price movement prediction. 

\paragraph{News} Investors often refer to social media and news, which significantly influence their investing behavior. News stories are the most easily accessible type of content on social media; therefore, we include information from news articles as one source of information in our analysis. We use retrieval methods to extract information from the news, summarize, and rank them based on relevance to provide to the model.

\paragraph{Company's Industry and Services/Products} The company's industry and services/products are important because different factors can affect one industry or one type of service/product. We include keywords of the company's industry and services as input for our analysis.

In our approach, we prompt pre-trained language models with the aforementioned information. We utilize a hypothetical user query to retrieve relevant information about a specific company at a specific time. For this purpose, we collect news articles and financial data for 20 highly traded stocks in the stock market. Our model and data will be available upon publication of this research paper.

\begin{table*}
    \small
    \centering
    \begin{tabular}{|p{16cm}|}
        \rowcolor{gray!30} \multicolumn{1}{c}
        {\bf{News Article Sample}} \vspace{1mm} \\[5pt]
        \textbf{Title:} Should You Invest in Apple (AAPL) Based on Bullish Wall Street Views? \\
        \textbf{Description:} Based on the average brokerage recommendation (ABR), Apple (AAPL)... \\
        \textbf{Data-Time:} April 12, 2024 at 6:30 AM \\
        \textbf{Keywords:} Strong Buy, Zacks Rank, brokerage firms, Apple,... \\
        \textbf{Content:} Investors often turn to recommendations made by Wall Street analysts... \\
        \textbf{URL:} https://news.google.com/articles/CBMiTWh0d... \\
        \textbf{Title:} Congress passes spending bill with TikTok ban on government devices \\
        \textbf{Description:} Congress passed a large spending package that includes a bill banning TikTok from being used on government devices and new filing fees for mergers. \\
        \textbf{Data-Time:} Dec 23, 2022 at 3:08 PM \\
        \textbf{Keywords:} Alphabet Class, Amazon, Meta, Apple, Joe Biden, Politics, Social media \\
        \textbf{Content:} \\
        \textbf{URL:} https://www.cnbc.com/2022/12/23/congress-passes-spending-bill-with-tiktok-ban-on-government-devices.htm \\ \\
        
        \rowcolor{gray!30}\multicolumn{1}{c}{\bf{Company Names (Tickers)}} \vspace{1mm} \\[5pt]
        Apple Inc. (AAPL), Amazon.com Inc. (AMZN), Alphabet Inc. (GOOGL), Microsoft Corporation (MSFT), Tesla, Inc. (TSLA), Facebook, Inc. (FB), Berkshire Hathaway Inc. (BRK.B), Johnson \& Johnson (JNJ), Visa Inc. (V), JPMorgan Chase \& Co. (JPM), Procter \& Gamble Company (PG), Walmart Inc. (WMT), The Coca-Cola Company (KO), Netflix Inc. (NFLX), Pfizer Inc. (PFE), Walt Disney Company (DIS), Nvidia Corporation (NVDA), Alibaba Group Holding Limited (BABA), Adobe Inc. (ADBE), Mastercard Incorporated (MA) \\ \\
        
        \rowcolor{gray!30}\multicolumn{1}{c}{\bf{News Websites}} \vspace{1mm} \\[5pt]
        Africa.businessinsider.com, Aljazeera.com, Apnews.com, Benzinga.com, Businessinsider.com, CNBC.com, CNN.com, CFO.economictimes.indiatimes.com, Decrypt.co, Defenseworld.net, Edition.cnn.com, Economist.com, Epaper.financialexpress.com, Etfdailynews.com, Financialbuzz.com, Fool.com, Forbes.com, Foxbusiness.com, FT.com, Globenewswire.com, Investingnews.com, Investorideas.com, Kiplinger.com, Markets.businessinsider.com, Moneycontrol.com, Newswire.ca, Pennystocks.com, Prnewswire.com, SCMP.com, Stockmarket.com, Stocknews.com, Theatlantic.com, Theweek.com, UPI.com, \\
        \midrule
    \end{tabular}
    \caption{Dataset Information}
    \label{tab:dataset}
\end{table*}\rule{0pt}{10pt}

\begin{table*}[t]
\small
    \centering
    \begin{tabular}{c|c|p{10cm}}
        \toprule
        \bf{Financial Variables} &  \textbf{Source} & \textbf{Description} \\
        \midrule
        Total Revenue&(10K) Reports&Aggregate amount of income generated from the sale of goods or services before deducting any expenses\\
        \rowcolor{lightgray!40}Net Income&(10K) Reports&Also known as profit or net earnings, is the total amount of revenue earned by a company after deducting all expenses\\
        Free Cash Flow&(10K) Reports&Represents the amount of cash generated by a company's operations\\
        \rowcolor{lightgray!40}Total Assets&(10K) Reports&Combined value of all resources owned or controlled by a company\\
        Price Momentum&Historical Pricing Data&Price Momentum 6-12 Months — Measures the relative strength and direction of a stock's price movement over the past months.\\
        \rowcolor{lightgray!40}Forward Return&Historical Pricing Data&Expected or projected return on an investment or asset over a future period. \greenlight{Used as target}\\
        \bottomrule
    \end{tabular}
    \vspace{2mm}
    \caption{Descriptions for the financial data available in financial reports extracted from Income Statements, balance sheets, and cash flow statements.}
    \label{tab:fin}
\end{table*}

\section{Literature Review}

\subsection{Prompting LLMs}

We can now prompt language models with a variety of tasks, and classification is a significant task that LLMs can solve. They are not only capable of classifying textual data but also able to classify tabular data, even in few-shot settings \cite{Fang2024-lr, Hegselmann2022-ng}. In finance, prompting language models have been used for tasks such as semantic analysis \cite{zhang2023enhancing, loukas2023making, vamvourellis2022learning}.

In addition to zero-shot prompting, we can enhance the performance of language models by providing a few examples of the task with the expected outcome before asking the main question \cite{Brown2020-ev}. This approach, known as few-shot prompting, does not involve a training process but has been proven to help the language model understand the task and the distribution of input and expected output.

Lewis et al. \cite{Lewis2020-ba}, by introducing retrieval-augmentation, showed that adding in-context information to the prompts can increase model accuracy and prevent hallucination. Kong et al. \cite{Kong2023-oe} and Pasunuru et al. \cite{Pasunuru2021-qy} proposed methods for keyphrase extraction and abstractive summarization for retrieval augmentation, respectively. We examined these methods and adapted their concepts to develop our own approach to perform retrieval augmentation.

\subsection{Stock Price Movement}

Traditional methods \cite{Weng2017-wl, Ou2009-kj} often utilize tabular classification formats with machine learning algorithms such as support vector machines, decision trees, and feedforward neural networks. For example, Weng et al. \cite{Weng2017-wl} employed a wide range of features, including the number of Google News articles about a company. Later studies shifted focus to forecasting and movement classification using sequential neural networks, such as LSTM and GRU, and some of them included CNN networks \cite{Song2023-lj, Nelson2017-hc, A2023-dl}  to the architecture.

Recent studies have explored using large language models to predict price movements \cite{Lopez-Lira2023-nv, Wang2024-cn, Tong2024-qm}. Wang et al. \cite{Wang2024-cn} introduced the LLMFactor model, which integrates historical price data with news articles, extracting key phrases, factors, and sentiments, and provided them in the prompt to enhance the analysis. Similarly, Tong et al. \cite{Tong2024-qm} combined news and historical price data, employing a strategy that leverages a range of ``experts'' (technical expert, semantic expert, and human expert) to analyze stock-related features. Their approach involves generating bullish and bearish rationales based on market conditions and expert opinions to make the final decision.

\section{Dataset}
 We collected data for 20 companies from 2021 to the present. This data includes financials collected from 10-K reports and news articles. The companies were chosen based on the volume of trading across different industries, so we selected companies whose stocks have been traded the most across different industries. List of news websites and companies, and a sample new article is available in Table \ref{tab:dataset}.

\subsection{News Articles}
We utilize a scraper to gather articles that discuss investment in the selected companies based on the tags/keywords of the news articles and the presence of the company's name in the title. We crawled the links using a simple search query on the news website mentioning specific dates and company names. Subsequently, we scraped the information from the generated articles and developed the dataset. A total of 5000 news articles were extracted across 20 companies from October 2021 to January 2024. A sample of our dataset is shown in Table \ref{tab:dataset}.

\subsection{Financial Data}

The financial data is gathered from the 10-K quarterly filings. These reports contain operational/income statements, cash flow statements, and balance sheets, which are available on the companies' websites. Table \ref{tab:fin} contains descriptions of some information found in the financial reports. 

Additionally we include historic price data in our study. We do not treat historical data as time-series data in this study. Instead, we capture price momentum by including changes in price over the past 6 and 12 months.

\section{Methodology}


\subsection{Retrieval Methods}

Due to the high number of news articles and the amount of information about each company on a given date, we need to find a method to extract the important information from the large length of articles. The filtering and summarizing process should be able to detect the relevant portions of the news articles and select the most relevant sections for the query.

\begin{figure*}
    \centering
    \includegraphics[width=500pt]{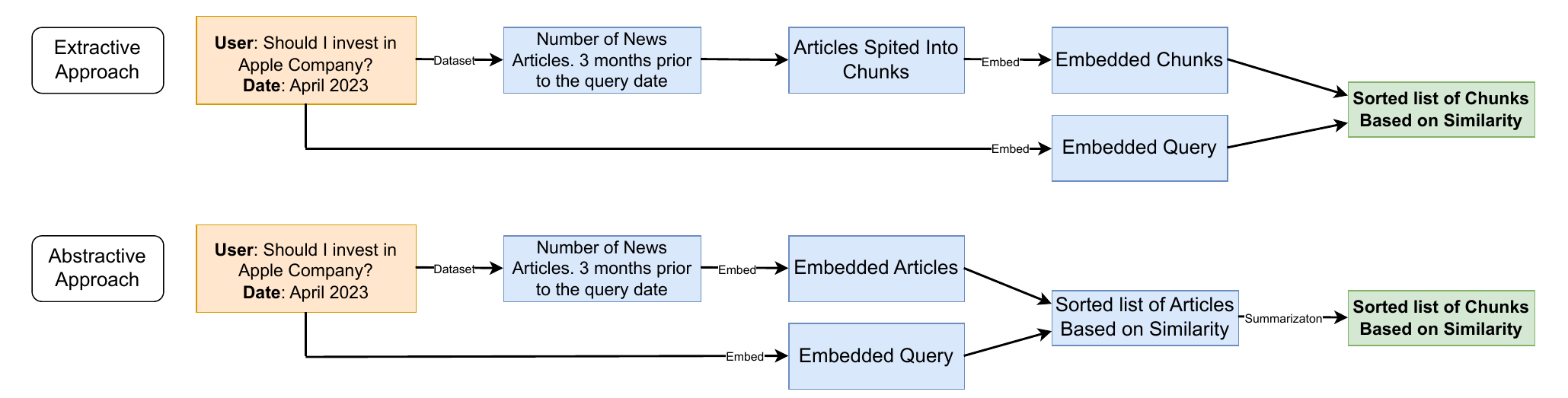}
    \caption{Summarizing flow diagram}
    \label{fig:flow}
\end{figure*}

Building upon the methodology outlined in Sarthi et al. \cite{Sarthi2024-rn}, we implemented a layered summarization approach. Initially, we filtered news articles based on key metadata such as title, subtitle, and publication date, searching for the company name and index ticker\footnote{A symbol, a unique combination of letters and numbers that represent a particular stock or security} within the articles. Subsequently, we proceeded with summarizing the selected articles. For comparison, we employed both of abstractive and extractive summarization techniques. For extractive summarization, we employed both OpenAI and Sentence-BERT embedding to encode chuncks of the news articles with three sentences in length and compute similarities with the user's query. Additionally, for abstractive summarization, we leveraged the GPT-3.5 model to generate summaries of the most relevant articles to the user's query. Figure \ref{fig:flow} shows the summarizing processes.

To simulate user's queries we tried a set of different queries such as:
\begin{itemize}
  \item Investing in \textit{company-name} company in \textit{date}.
  \item Should I invest in \textit{company-name} in company \textit{date}
  \item Is \textit{company-name} company index bullish/going up after \textit{date} or bearish/going down?
\end{itemize}

Table \ref{tab:sumarizations} shows summaries examples. The chunks of articles can be from one or multiple news articles. The two methods, OpenAI and Sentence BERT embeddings, can result in the same chunks. Moreover, one method can recommend multiple chunks from the same article.

Based on our observations, the granularity of chunks can affect information extraction. Lower chunk sizes can result in abstract "Invest/Do Not Invest" summaries, while larger chunk sizes can provide more explanations. Similarly, extractive summaries can highlight parts of the text mentioning stock price movements, while abstractive summarization can include other information from the text as well. Finally, we decided to proceed with extractive summarization with a chunk size of three sentences using OpenAI embedding to retrieve more pertinent information.

\begin{figure*}
    \centering
    \includegraphics[width=500pt]{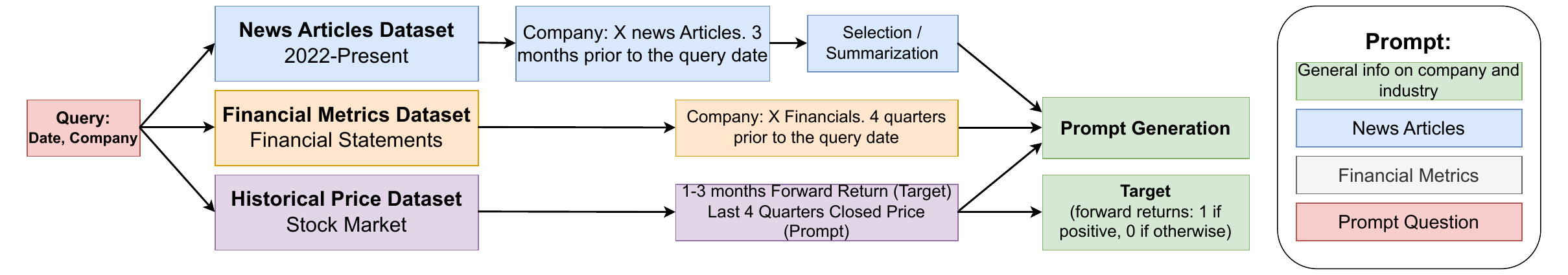}
    \caption{Prompt generation flow diagram.}
    \label{fig:prompt}
\end{figure*}

\subsection{Prompt Generation and Labeling}

Prompt format plays a crucial role in the performance of  models, as the structure and order of information can significantly impact the model's ability to generate accurate responses. We experimented with different orders of information and prompt placements, such as placing the prompt question at the beginning or the end, and tried various prompt questions until we found the one that yielded the best results.

After collecting news articles and financial data, our next step involved generating the prompts. Each prompt contains a sentence of the company's products and its industry, up to 6 chunks of the most relevant recent news articles, financial features for the previous four quarters, and finally, the prompt question. The labels for each prompt would be the forward return in binary format for upcoming three and six months after the date mentioned in the prompt question. Our prompts typically average around 2,500 tokens in length, as they incorporate information from multiple sources. A sample promt is shown in Tabel \ref{tab:sources}

For instance, if the prompt question is: ``Should I invest in Apple Company in July 2022?'' The news articles will be sampled from May-June 2022, and the financial features corresponding the last four quarters before June 2022 will be included. Furthermore, we will assess the target based on whether the stock in question experienced an increase (labeled as 1) or decrease (labeled as 0) in price in the months following (referred to as forward return in Table \ref{tab:fin}).

In our analysis, we examined the language model's responses across three, and six months of forward return to determine the optimal time period for prediction. The general prompt generation process is visualized in Figure \ref{fig:prompt}.

\paragraph{Company's General Information}

We included a brief statement about the company's products, primary activities, and the industry it operates within. This addition serves two purposes. Firstly, stock prices are often influenced by industry trends, and we want the language model to consider this context when generating responses. Secondly, since the prompts are situated in the past (e.g., a prompt from February 2022), the target outcome has already occurred. This introduces a risk of data leakage, as the language model has already been trained on information available online. Therefore, in the prompt generation process, we replace the company's name with a placeholder like ``Company.'' However, to provide necessary context about the company's operations, we include a general information sentence at the beginning of the prompt. 



\begin{table*}
    \small
    \renewcommand{\arraystretch}{1.2}
    \centering
    \begin{tabular}{p{16cm}}
    
        \rowcolor{gray!30} \multicolumn{1}{c}{\bf{Sample Prompt}}\\[5pt]
        \rowcolor{green!15} \textbf{General info on company and industry:} \\
        \rowcolor{green!10} Amazon is a leader in the e-commerce and cloud computing sectors, pioneering new standards in online retail and services. \\
        \rowcolor{cyan!15}\textbf{Recent news about Amazon:} \\
        \rowcolor{cyan!10}\textbf{Title:} Better Buy: Amazon or All 19,941 Cryptocurrencies? | The Motley Fool \\
        \rowcolor{cyan!10}\textbf{Summary:} If I had to choose between owning all of the existing crypto projects or Amazon stock, I think I'd go with Amazon. The sell-off in the stock presents an attractive buying opportunity. The company's low-margin e-commerce business is struggling amid persistent supply chain bottlenecks, higher shipping costs, and intensifying competition. \\
        \rowcolor{cyan!10}\textbf{Title:} Better Buy: Amazon or All 19,941 Cryptocurrencies? | The Motley Fool \\
        \rowcolor{cyan!10}\textbf{Summary:} That puts the value of the entire crypto market just below that of Amazon (AMZN -1.14\%), which despite being down over 40\% from its high is still worth more than \$1 trillion. Given their similar total valuations, it seems natural to compare the two and ask which would be the better investment now: Amazon or a basket of all 19,941 crypto tokens? \\
        \rowcolor{gray!15}\textbf{Last four quarters financial information for Amazon:} \\
        \rowcolor{gray!10}\textbf{Date:} June 2022, \textbf{Total Revenue:} \$121.23B, \textbf{Net Income:} -\$2.03B, \textbf{EPS:} -0.2, \textbf{Free Cash Flow:} -\$6.76B, \textbf{Total Assets:} \$419.73B, \textbf{Close Price:} \$106.21 \\
        \rowcolor{gray!10}\textbf{Date:} March 2022, \textbf{Total Revenue:} \$116.44B, \textbf{Net Income:} -\$3.84B, \textbf{EPS:} -0.378, \textbf{Free Cash Flow:} -\$17.74B, \textbf{Total Assets:} \$410.77B, \textbf{Close Price:} \$162.99 \\
        \rowcolor{gray!10}\textbf{Date:} December 2021, \textbf{Total Revenue:} \$137.41B, \textbf{Net Income:} \$14.32B, \textbf{EPS:} 1.411, \textbf{Free Cash Flow:} \$3.15B, \textbf{Total Assets:} \$420.55B, \textbf{Close Price:} \$166.72 \\
        \rowcolor{gray!10}\textbf{Date:} September 2021, \textbf{Total Revenue:} \$110.81B, \textbf{Net Income:} \$3.16B, \textbf{EPS:} 0.312, \textbf{Free Cash Flow:} -\$8.44B, \textbf{Total Assets:} \$382.41B, \textbf{Close Price:} \$164.251 \\
        \rowcolor{red!10} \textbf{Question:} Is the price for COMPANYX going UP or DOWN, binary classify on [UP] or [DOWN]. [UP] if news and financials are predicting increase in price and bullish market, [DOWN] if news and financials are predicting decrease in price and bearish in next 3 months if [UP] percentage should be positive, if [DOWN], percentage should be negative." \\
    \end{tabular}
    \vspace{2mm}
    \caption{Sample Prompt; Before Prompting the LLMs, the name of the company, in this case `Amazon', will be replaced by an unknown compnay name such as Company-X}
    \label{tab:sources}
\end{table*}

\subsection{Models}

We use a wide range of pre-trained language models. We utilized Meta's LLaMA2 7B, 13B, and 70B versions, LLaMA3 8B and 70B versions, and OpenAI's GPT-3.5 and GPT-4 models.

Noteworthy is the fact that some prompts in four-shot learning were so large that they exceeded the maximum input token limit of LLaMA2 models. As a result, we excluded LLaMA2 from the four-shot analysis.

\section{Experimental Setup}

The prompting process was done in zero, two, and four-shot formats. For zero-shot learning, we presented the language model with a set of prompts containing the user query, the company information, a set of generated text articles, and the corresponding financial metrics. In order to test the prediction accuracy of our model, we prepared a set of 120 sample prompts. We then repeated the same process in two and four-shot settings, where we presented the language model with two and four example prompts with answers (``[DOWN]'' if the stock price decreases and ``[UP]'' if it increases). We selected the samples so that there is one of each class in the two-shot examples, and two of each class in the four-shot examples, so the language model can be familiar with the answer distribution.




\begin{table*}[t]
\centering
    \resizebox{2.0\columnwidth}!{
    \begin{tabular}{c|c|*{7}{c}|*{7}{c}}
        \toprule
        \multicolumn{2}{c}{\bf{}} & \multicolumn{7}{|c}{\bf{3 Months Forward Return}} & \multicolumn{7}{|c}{\bf{6 Months Forward Return}}\\
        \midrule
        \bf{} & \bf{Model} & NP & PP & NR & PR & ACC & MCC & WF1 & NP & PP & NR & PR & ACC & MCC & WF1 \\ 
        \midrule
        \multirow{7}{*}{\centering \bf{Zero-Shot}} 
        & \bf{LLaMA2 7B} &0.200&0.565&0.008&0.997&0.565&0.014&$0.413_{0.016}$&0.267&0.561&0.012&0.977&0.556&-0.029&$0.412_{0.019}$\\  
        & \bf{LLaMA2 13B} &0.367&0.557&0.062&0.913&0.542&-0.088&$0.436_{0.033}$&0.369&0.559&0.058&0.923&0.545&-0.054&$0.436_{0.019}$\\ 
        & \bf{LLaMA2 70B} &0.410&0.555&0.217&0.758&0.522&-0.030&$0.484_{0.014}$&0.355&0.538&0.183&0.739&0.496&-0.073&$0.455_{0.028}$\\ 
        & \bf{LLaMA3 8B} &0.535&0.613&0.408&0.726&0.587&\inc{20}{0.143}&\dec{20}{$0.577_{0.012}$}&0.539&0.614&0.404&0.732&0.589&\inc{80}{0.143}&\dec{90}{$0.578_{0.003}$}\\ 
        & \bf{LLaMA3 70B} &0.533&0.600&0.333&0.774&0.582&0.120&$0.560_{0.000}$&0.533&0.600&0.333&0.774&0.582&\inc{20}{0.120}&\dec{20}{$0.560_{0.000}$}\\ 
        & \bf{GPT3.5} &0.563&0.623&0.412&0.752&0.604&\inc{80}{0.175}&\dec{90}{$0.592_{0.016}$}&0.504&0.591&0.333&0.745&0.565&0.086&$0.546_{0.020}$\\ 
        & \bf{GPT4}&0.531&0.622&0.462&0.684&0.587&\inc{50}{0.150}&\dec{50}{$0.583_{0.009}$}&0.519&0.615&0.454&0.674&0.578&\inc{50}{0.131}&\dec{50}{$0.574_{0.012}$}\\ 
        \midrule
        \multirow{7}{*}{\centering \bf{Two-Shot}} 
        & \bf{LLaMA2 7B} &0.436&0.400&0.992&0.006&0.436&-0.010&$0.271_{0.011}$
        &0.431&0.394&0.950&0.029&0.431&-0.054&$0.289_{0.019}$\\  
        & \bf{LLaMA2 13B} & 0.443&0.569&0.412&0.600&0.518&-0.088&$0.514_{0.024}$&0.416&0.559&0.200&0.784&0.529&-0.020&$0.485_{0.021}$\\ 
        & \bf{LLaMA2 70B} &0.489&0.595&0.408&0.671&0.556&0.081&$0.549_{0.031}$&0.448&0.574&0.542&0.481&0.507&0.022&$0.508_{0.024}$\\ 
        & \bf{LLaMA3 8B} &0.518&0.621&0.492&0.645&0.578&\inc{20}{0.138}&\dec{20}{$0.577_{0.012}$}&0.452&0.580&0.529&0.503&0.515&0.032&\dec{20}{$0.516_{0.009}$}\\
        & \bf{LLaMA3 70B} &0.524&0.584&0.229&0.839&0.573&0.086&$0.527_{0.000}$&0.490&0.577&0.221&0.823&0.560&\inc{20}{0.054}&$0.515_{0.006}$\\ 
        & \bf{GPT3.5} &0.529&0.620&0.462&0.681&0.585&\inc{50}{0.146}&\dec{50}{$0.581_{0.027}$}&0.497&0.590&0.346&0.729&0.562&\inc{50}{0.081}&\dec{50}{$0.545_{0.019}$}\\ 
        & \bf{GPT4} &0.521&0.640&0.567&0.597&0.584&\inc{80}{0.162}&\dec{90}{$0.585_{0.012}$}&0.526&0.647&0.579&0.597&0.589&\inc{80}{0.175}&\dec{90}{$0.591_{0.011}$}\\ 
        \midrule
        \multirow{4}{*}{\centering \bf{Four-Shot}} 
        & \bf{LLaMA3 8B} &0.482&0.597&0.465&0.613&0.548&\inc{20}{0.078}&\dec{50}{$0.547_{0.018}$}&0.463&0.643&0.792&0.290&0.509&\inc{50}{0.094}&$0.481_{0.006}$\\
        & \bf{LLaMA3 70B} &0.476&0.573&0.208&0.823&0.555&0.039&$0.507_{0.000}$&0.522&0.586&0.250&0.823&0.573&\inc{20}{0.089}&\dec{50}{$0.533_{0.000}$}\\ 
        & \bf{GPT3.5} &0.505&0.591&0.326&0.753&0.567&\inc{50}0.087&\dec{20}{$0.546_{0.013}$}&0.455&0.569&0.215&0.801&0.545&0.020&\dec{20}{$0.502_{0.016}$}\\ 
        & \bf{GPT4} &0.494&0.619&0.556&0.559&0.558&\inc{80}{0.114}&\dec{90}{$0.559_{0.004}$}&0.514&0.624&0.514&0.624&0.576&\inc{80}{0.138}&\dec{90}{$0.576_{0.019}$}\\ 
        \bottomrule
    \end{tabular}
    }
    \vspace{2mm}
    \caption{Negative precision (NP), positive precision (PP), negative recall (NR), positive recall (PR), weighted F1-score (WF1), Matthews correlation coefficient (MCC), and accuracy (ACC) metrics for stock price movement binary classification are shown in this table. Five LLaMA models and two GPT models are used to predict the movement of the price in the next 3 and 6 months. \decrease and \increase highlights show the best models for each group of N-shot learning of the LLMs based on WF1 and MCC respectively. The results are the average of 10 runs, and WF1 sub-texts show the standard deviation of WF1 among 10 runs.}
    \label{tab:results}
\end{table*}

\begin{figure*}
    \centering
    \includegraphics[width=450pt]{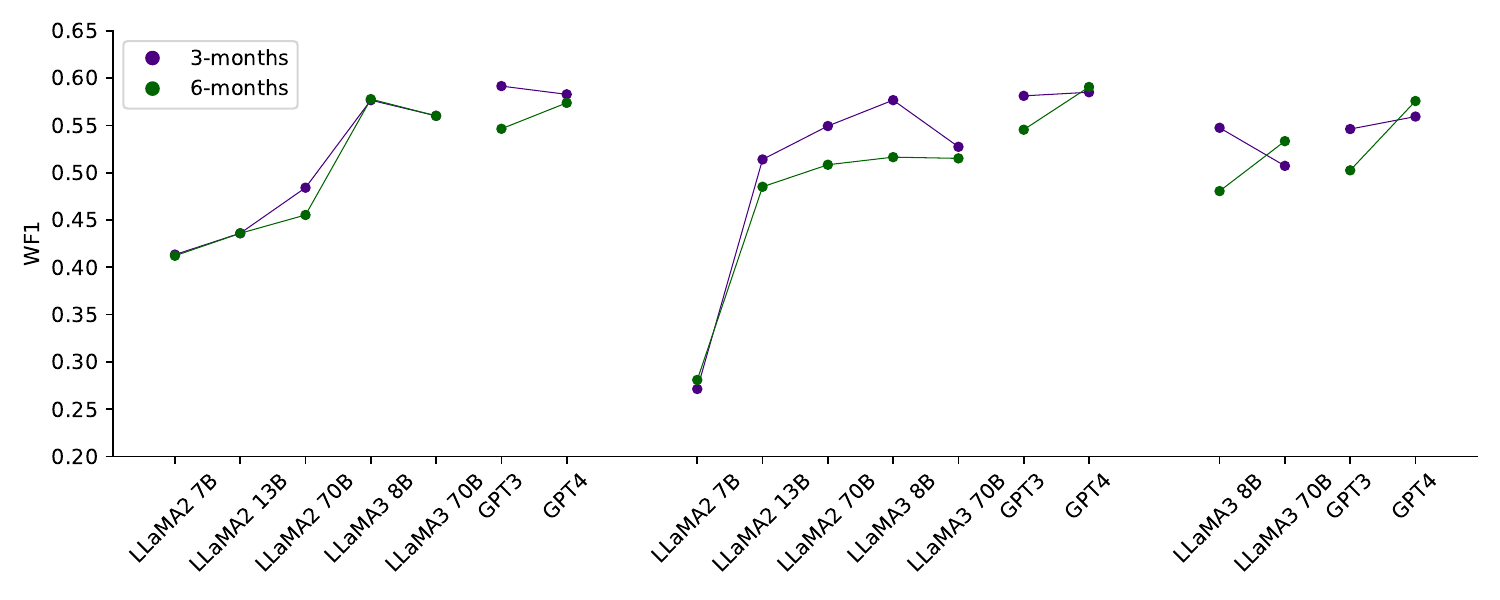}
    \caption{Weighted F1 scores for single stock price movement binary classification. The results are show for both 3 and 6 months prediction for LLaMA and GPT models.}
    \label{fig:wf1}
\end{figure*}

\subsection{Evaluation Metrics}

Initially we used accuracy defined as $\frac{\text{Correct Predictions}}{\text{Total Predictions}}$ reflects the overall correctness of the model, but cannot capture its performance in cases where the dataset is imbalanced.

56.4 and 70.1 percent of our test samples are stocks that appreciate in price in the next three and six months respectively. Metrics such as precision and recall can capture the performance of the model in each class. Precision measures the accuracy of each class's prediction, while recall measures the completeness of predictions.

Given the confusion matrix 
\(\left[\begin{matrix}
\text{tp} & \text{fp} \\
\text{fn} & \text{tn}
\end{matrix}\right]\) we define our metrics:

\[
\text{Precision - positive class (PP)} = \frac{TP}{TP + FP}
\]

\[
\text{Recall - positive class (PR)} = \frac{TP}{TP + FN}
\]

F1-scores evaluates the model's ability to balance precision and recall for both positive and negative classes. To make F1-score more general and take into account the number of samples in each class in the test set, we use the Weighted F1-score defined as: 

$$
\text{Weighted F1-Score (WF1)} = \frac{\sum_{i=1}^{C} w_i \times F1_i}{\sum_{i=1}^{C} w_i}
$$

where \(C\) is the number of classes, \(w_i\) is the weight for class \(i\), and \(F1_i\) is the F1-score for class \(i\). The WF1 provides a balanced assessment that accounts for imbalances in class distribution within the test set.

The Matthews Correlation Coefficient (MCC) is a metric used to evaluate the performance of a classification model, particularly in imbalanced datasets. It takes into account true and false positives and negatives, providing a more informative measure than accuracy. The MCC ranges from -1 to 1 where 1 indicates perfect prediction, 0 corresponds to random chance and  -1 represents total disagreement between predicted  and actual classification. The formula for MCC is given by:

\[
\text{MCC} = \frac{{TP \times TN - FP \times FN}}{{\sqrt{(TP + FP)(TP + FN)(TN + FP)(TN + FN)}}}
\]

The MCC and WF1 metrics do not always align with each other. While our primary focus is on the WF1 score, we also discuss the MCC results in our analysis for a comprehensive evaluation.

\subsection{Statistical Significance of The Results}

Since LLMs are generative models, we prompted the language models listed in Section 4.3 using their respective APIs and repeated our evaluations five times to ensure the statistical significance of the results. Finally, we computed the standard deviation for the WF1 scores to assess the consistency of the model performance across the repeated trials.

\section{Results and Discussions}

In our results and discussions, we primarily focus on the weighted F1 score (WF1) as it provides a balanced metric for evaluating both classes. The results are shown in table \ref{tab:results} and WF1 score are shown in Figure \ref{fig:wf1}. The best-performing models for both 3-month and 6-month predictions were the GPT and LLaMA3-8B models. We observed that adding more examples in two-shot and four-shot learining did not significantly improve the performance of these top models.

\paragraph{Effect of the Language Models}
The results in Figure \ref{fig:wf1} demonstrate that more advanced models, such as LLaMA3 and GPT models, outperform LLaMA2 models. Additionally, LLaMA2 models show improved performance with an increase in the number of parameters. GPT-4 models generally exhibit better performance than GPT-3.5 models based in both WF1 and MCC, except in the 3-month predictions with zero-shot setting that GPT3.5 has a better WF1 and MCC. 

An interesting observation is that the performance of LLaMA3 models declines based on WF1 when increasing the number of parameters from 8B to 70B, a trend consistent in five out of six cases; This observation is also consistent based on MCC. Finally, GPT-4 models and LLaMA3-8B models WF1 scores are generally in the same range.

Specifically, the highest WF1 score for the 3-month prediction was 0.592, achieved by GPT-3.5 with zero-shot setting. For the 6-month prediction, GPT-4 with two-shot learning achieved the best result, with a WF1 score of 0.591 percent.

Based on MCC results, the best models are stil LLaMA3 and GPT models. However, the best performances for 3 month movement prediction is GPT3.5 with zero-shot setting (0.175) and GPT4 with two-shot setting (0.162). For 6 months movement prediction, the best performing model is GPT4 with two-shot learning (0.175).

\paragraph{Effects of Few-Shot Learning}
The results presented in Table \ref{tab:results} and Figure \ref{fig:wf1} indicate that the performance of the best models does not improve significantly with regard to WF1 scores and accuracies. This may be due to the length of each prompt, which averages 2,500 tokens, resulting in very long prompts when stacking two or four examples. Such long prompts can confuse the language model, particularly smaller versions like LLaMA2 7B, which showed a decrease in performance after adding two examples. However, LLaMA2 13B and 70B models experienced improved performance with the addition of few-shot learning. 

For top-performing models like LLaMA3 and GPT, LLaMA3 models generally saw a decrease in performance with an increased number of examples, except for LLaMA3-70B, which achieved 0.576 WF1 score in the four-shot setting for 6-month predictions. GPT models exhibited a similar trend, though GPT-4 for 6-month forward returns showed an improvement from zero to two-shot learning (0.591 WF1).

The increase in performance from zero-shot to two-shot setting, and decrease in performance from two-shot to four-shot setting is detectable by MCC results. Generally two-shot models are the highest performing models.

\paragraph{Standard Deviations}
The standard deviations of WF1 scores within the 10 runs range between 0 and 0.033, indicating that the results are statistically stable. Less advanced models, such as the LLaMA2 variants, generally exhibit higher standard deviations.

\paragraph{Prediction Intervals}
The general trend indicates that 3-month predictions typically show better performance compared to 6-month predictions according to both MCC and WF1 score. However, increasing the number of few-shot examples appears to improve predictions for models like LLaMA3-8B and GPT-4, making their 6-month forecasts more accurate than their 3-month based on WF1 score.

\section{Future Studies}

Our initial goal was to leverage the language model in a few-shot setting. We have observed that increasing the number of examples in few-shot setting may help the language model become familiar with the output distribution but does not necessarily enhance performance for this specific task. Therefore, we plan to shift our focus to fine-tuning smaller language models with a combination of textual and tabular data.

Additionally, predicting the percentage of change in forward return in a regression setting instead of relying on binary classification for price movement will be our next area of focus.


\begin{table*}
    \small
    \centering
    \begin{tabular}{c|p{14cm}}
        \toprule
        \bf{Query Info} & \textbf{Query:}
        Should I invest in Visa company Feb 2024?
        
        \textbf{Count:} 3
        
        \textbf{Chunk Size:} 3\\

        \midrule
        \bf{Extractive: OpenAI}& 
        
        \textbf{Chunk 1:} similarity: 0.660
        
        \textbf{Article Title:} Visa vs. Mastercard: Which Stock Is the Better Buy Today?
        \textbf{Text:}If you want to buy into blue chip businesses with long runways for sustainable growth, \greenlight{consider buying both.} Should you invest \$1,000 in Visa right now? Before you buy stock in Visa, consider this: The Motley Fool Stock Advisor analyst team just identified what \greenlight{they believe are the best stocks for investors to buy now… and Visa wasn’t one of them.}
        
        \textbf{Chunk 2:} similarity: 0.659
        
        \textbf{Article Title:} Want \$1 Million in Retirement? Invest \$50,000 in These 3 Stocks and Wait a Decade.
        \textbf{Text:} Should you invest \$1,000 in Visa right now? Before you buy stock in Visa, consider this: The Motley Fool Stock Advisor analyst team just identified what they believe are the \greenlight{best stocks for investors to buy now… and Visa wasn’t one of them.} The 10 stocks that made the cut could produce monster returns in the coming years.
        
        \textbf{Chunk 3:} similarity: 0.651
        
        \textbf{Article Title:} The Zacks Analyst Blog Highlights NVIDIA, Visa, Amgen, Chubb and PACCAR.
        \textbf{Text:} The company's strategic acquisitions and alliances are fostering long-term growth and consistently driving its revenues. It expects net revenues to increase in the low double digits for fiscal 2024. \greenlight{Visa, fueled by increased payments and sustained investments in technology, is witnessing bottom-line growth.} The ongoing shift to digital payments is advantageous for Visa, with strong domestic volumes supporting its overall performance.
        \\
        
        \midrule
        \bf{Extractive:S-BERT}&
        \textbf{Chunk 1:} similarity: 0.724
        
        \textbf{Article Title:} Want \$1 Million in Retirement? Invest \$50,000 in These 3 Stocks and Wait a Decade.
        \textbf{Text:} Should you invest \$1,000 in Visa right now? Before you buy stock in Visa, consider this: The Motley Fool Stock Advisor analyst team just identified what \greenlight{they believe are the best stocks for investors to buy now… and Visa wasn’t one of them.} The 10 stocks that made the cut could produce monster returns in the coming years.

        \textbf{Chunk 2:} similarity: 0.672
        
        \textbf{Article Title:} The Zacks Analyst Blog Highlights NVIDIA, Visa, Amgen, Chubb and PACCAR
        \textbf{Text:} The company's strategic acquisitions and alliances are fostering long-term growth and consistently driving its revenues. It expects net revenues to increase in the low double digits for fiscal 2024. Visa, fueled by increased payments and sustained investments in technology, is witnessing bottom-line growth. The ongoing shift to digital payments is advantageous for Visa, with strong domestic volumes supporting its overall performance.

        \textbf{Chunk 3:} similarity: 0.628
        
        \textbf{Article Title:} Want \$1 Million in Retirement? Invest \$50,000 in These 3 Stocks and Wait a Decade.
        \textbf{Text:} \greenlight{At any rate, Visa's no-nonsense business model has delivered exceptional shareholder gains.} Over the last decade, Visa's stock has generated an annualized total return (price appreciation plus dividend payments) of 18.4\%. That means \$50,000 invested in Visa 10 years ago would be worth \$271,000 today -- not too shabby.
        \\
        
        \midrule
        \bf{Abstractive: GPT-3.5}&
        \textbf{Chunk 1:} similarity: 0.676
        
        \textbf{Article Title:} Why Visa (V) is a Top Stock for the Long-Term. 
        \textbf{Text:} \greenlight{Warren Buffett's Berkshire Hathaway holds a massive portfolio of stocks, including American Express, Visa, and Mastercard.} These three companies dominate the payments landscape and have strong economic moats, making them attractive investments for long-term investors. While their valuations may seem high, their profitable business models and network effects make them unlikely to be disrupted. Despite not being top picks for Buffett currently, these stocks should be on investors' radar for potential long-term growth.
        
        \textbf{Chunk 2:} similarity: 0.627
        
        \textbf{Article Title:} Visa (V) Enhances Offerings With Virtual Corporate Cards.
        \textbf{Text:} The article discusses the potential for cryptocurrency, specifically Bitcoin, to become the 12th sector of the S\&P 500 in the next few years. This is due to factors such as widespread adoption, the stability of stablecoins, and the success of companies like MicroStrategy who have heavily invested in Bitcoin. The recent approval of Bitcoin ETFs by the U.S. Securities and Exchange Commission has also opened the door for institutional adoption. The article suggests that with a current market cap of over \$2 trillion, the cryptocurrency market has a chance to become a major player in the stock market.
        
        \textbf{Chunk 3:} similarity: 0.584
        
        \textbf{Article Title:} Visa vs. Mastercard: Which Stock Is the Better Buy Today?
        \textbf{Text:} Building an investment portfolio from scratch can be overwhelming, but \greenlight{focusing on stocks that are set to outperform the market, such as Visa (V), can lead to success.} V has been on the Zacks Focus List since 2017 and has seen a 191.34\% increase in shares since then. With analysts expecting earnings to grow 13\% for the current fiscal year, buying stocks with rising earnings estimates like V can be profitable.
        \\
        \bottomrule
        
    \end{tabular}
    \vspace{2mm}
    \caption{Comparison on summarization methods. \greenlight{The purple sentences} are the ones that seem explicitly informative in user's eyes. The similarity values are referring to cosine similarity between chunks and query.}
    \label{tab:sumarizations}
\end{table*}

\newpage
\clearpage

\bibliographystyle{IEEEtran}
\bibliography{main}

\end{document}